\documentclass{ws-procs961x669}            
\begin{document}
\title{A tale of two double quasars:\\
Hubble constant tension or biases?}

\author{L. J. Goicoechea$^*$ and V. N. Shalyapin$^{**}$}

\address{GLENDAMA Project Core Team, Universidad de Cantabria,\\
Avda. de Los Castros 48, E-39005 Santander, Spain\\
gravlens.unican.es\\
$^*$E-mail: goicol@unican.es\\
$^{**}$Main addresses: O.Ya. Usikov Institute for Radiophysics and Electronics,\\ 
National Academy of Sciences of Ukraine,\\
12 Acad. Proscury St., UA-61085 Kharkiv, Ukraine\\
and\\
Institute of Astronomy of V.N. Karazin Kharkiv National University,\\ 
Svobody Sq. 4, UA-61022 Kharkiv, Ukraine}

\begin{abstract}
For a flat $\Lambda$CDM (standard) cosmology, a small sample of gravitationally lensed 
quasars with measured time delays has recently provided a value of the Hubble constant $H_0$
in tension with the {\it Planck\/} flat $\Lambda$CDM result. Trying to check if this tension 
is real or not, we used basic observational constraints for two double quasars of the 
GLENDAMA sample (SBS 0909+532 and SDSS J1339+1310) to discuss the underlying value of $H_0$
in a standard cosmology. For SBS 0909+532, we were not able to obtain a reliable measurement 
of $H_0$. However, the current data of SDSS J1339+1310 are consistent with $H_0$ around 67.8 
km s$^{-1}$ Mpc$^{-1}$ and $\sigma (H_0)/H_0 \sim$ 10\%. Although the formal uncertainty is 
still large and mainly due to the lack of details on the mass density profile of the main 
lens galaxy, the central value of $H_0$ coincides with that of the TDCOSMO+SLACS  
collaboration (using gravitational lens systems) and is within the 1$\sigma$ interval from
{\it Planck\/} cosmic microwave background data. After getting these preliminary encouraging 
results through only one double quasar, we are currently planning to use several GLENDAMA 
systems to accurately measure the Hubble constant and put constraints on other cosmological 
parameters.
\end{abstract}


\bodymatter

\section{Introduction}\label{goico:sec1}
Optical photometric monitoring of gravitationally lensed quasars (GLQs) brings plenty of 
astrophysical information\cite{schn06}. For example, time delays between correlated brightness
variations of their multiple images are used to estimate the current expansion rate of the 
Universe (the so-called Hubble constant $H_0$), provided lensing mass distributions can be 
constrained by observational data\cite{jack15,treu16}. Throughout this paper, $H_0$ is expressed 
in standard units of km s$^{-1}$ Mpc$^{-1}$, so units only are explicitly given in tables and 
figures. 

Very recently, the H0LiCOW collaboration performed a joint analysis of six GLQs with measured 
time delays\cite{wong20}. For a flat $\Lambda$CDM standard cosmology, they obtained $H_0$ = 
73.3$^{+1.7}_{-1.8}$, in good agreement with $H_0$ = 74.03 $\pm$ 1.42 from SNe data by the SH0ES 
collaboration\cite{ries19}, but in apparent tension with Cosmic Microwave Background (CMB) 
data\cite{plan20}. {\it Planck\/} observations of the CMB suggested a narrow 1$\sigma$ interval 
ranging from 66.9 to 67.9, which is clearly inconsistent with H0LiCOW/SH0ES results. It is also 
worth noting that Freedman {\it et al.\/}\cite{free19} obtained an intermediate value of $H_0$ = 69.8 
$\pm$ 1.9.

The big question is whether the tension between early and late-Universe probes is due to 
systematic errors or has a physical origin. Possible systematic errors in some methods may fix 
this issue, avoiding hasty rejection of the standard cosmological model. Thus, we use two doubly
imaged quasars of the Gravitational LENses and DArk MAtter (GLENDAMA) sample\cite{gilm18} to 
discuss the influence of observational constraints, and hypotheses and priors on the mass model 
in the estimation of the Hubble constant in a standard cosmology. \Sref{goico:sec2} briefly 
presents the GLENDAMA project and the framework of time-delay cosmography through double 
quasars, while \sref{goico:sec3} and \sref{goico:sec4} include preliminary results for the GLQs 
SBS 0909+532 and SDSS J1339+1310, respectively. A discussion of results and future prospects 
appear in \sref{goico:sec5}. 

\section{GLENDAMA project and $H_0$ from doubles}\label{goico:sec2}
\begin{figure}[h]
\begin{center}
\includegraphics[width=0.7\textwidth]{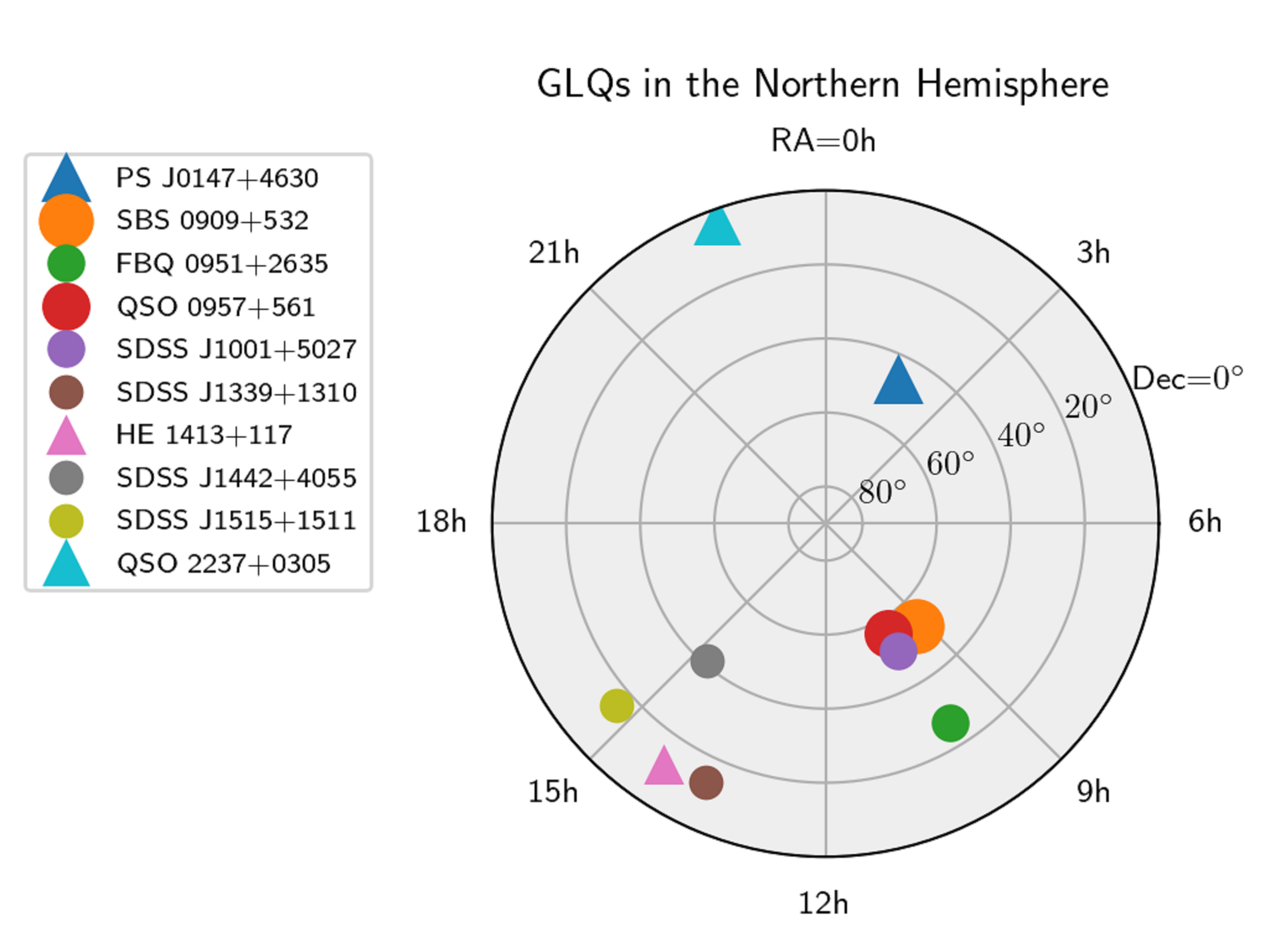}
\end{center}
\caption{GLENDAMA GLQs in the Northern Hemisphere. Triangles and circles represent 
quadruply and doubly imaged quasars, respectively. Larger symbols mean brighter quasars.}
\label{goico:fig1}
\end{figure}

The GLENDAMA project\footnote{\url{https://gravlens.unican.es/}.} is aimed to accurately study a 
sample of ten GLQs in the Northern Hemisphere over a period of about 25 years, basically 
covering the first quarter of this century\cite{gilm18}. The sample includes seven double 
quasars with two images (A and B) each, and three quads having four images (A, B, C and D) each.
\Fref{goico:fig1} shows the distribution on the sky of the selected, optically bright GLQs. The 
Gran Telescopio CANARIAS (GTC) is being used to obtain deep spectroscopy of the lens systems, 
while the optical variability of quasar images is traced from light curves mainly based on 
observations with the Liverpool Telescope (LT). These optical light curves are allowing us to 
measure the time delay $\Delta t_{\rm AB}$ in doubles, and three independent delays $\Delta 
t_{\rm AB}$, $\Delta t_{\rm AC}$ and $\Delta t_{\rm AD}$ in quads (see results in 
\tref{goico:tbl1}). Current GLENDAMA delays have been estimated to a typical accuracy of about 
5\%, with only two exceptions: the relative error in the delay of QSO 0957+561 is well below 
5\%, and the three delays of the quad HE 1413+117 have larger relative uncertainties.

\begin{table}
\tbl{GLENDAMA time delays from light curves in the SDSS $r$ band.}
{\begin{tabular}{@{}lcccl@{}}
\toprule
GLQ & $\Delta t_{\rm AB}$ & $\Delta t_{\rm AC}$ & $\Delta t_{\rm AD}$ & Reference \\
& (days) & (days) & (days) & \\
\colrule
PS J0147+4630   &   pm            &   pm       &  pm        & --- \\ 
SBS 0909+532    & 50$^{+2}_{-4}$  &  ---       &  ---       & Ref.~\citenum{hain13} \\
FBQ 0951+2635   &   pm            &  ---       &  ---       & --- \\
QSO 0957+561    & 420.6 $\pm$ 1.9 &  ---       &  ---       & Ref.~\citenum{shal12} \\
SDSS J1339+1310 & 48 $\pm$ 2      &  ---       &  ---       & Ref.~\citenum{shal21} \\
HE 1413+117     & 17 $\pm$ 3      & 20 $\pm$ 4 & 23 $\pm$ 4 & Ref.~\citenum{goic10} \\
SDSS J1442+4055 & 25.0 $\pm$ 1.5  &  ---       &  ---       & Ref.~\citenum{shal19} \\
SDSS J1515+1511 & 211 $\pm$ 5     &  ---       &  ---       & Ref.~\citenum{shal17} \\ 
\botrule
\end{tabular}
}
\begin{tabnote}
pm = preliminary measure.\\
\end{tabnote}
\label{goico:tbl1}
\end{table}

The time delay between the two images of a double quasar can be expressed in terms of the 
so-called time-delay distance $D_{\Delta t}$, the speed of light $c$ and a dimensionless factor 
$\Delta \Phi_{\rm AB}$, so that\cite{treu16} $\Delta t_{\rm AB} = (D_{\Delta t}/c) \Delta 
\Phi_{\rm AB}$. Here, $D_{\Delta t}$ depends on the source (quasar) and deflector (main lens 
galaxy) redshifts, as well as cosmological parameters. Measuring the redshifts, and assuming a 
flat $\Lambda$CDM cosmological model with $\Omega_{\rm M}$ = 0.3 (matter density) and 
$\Omega_{\Lambda}$ = 0.7 (dark energy density), $D_{\Delta t}/c$ is given as a known constant 
divided by $H_0$\footnote{The time-delay distance does not appreciably change when matter and 
dark energy densities are slightly different to 0.3 and 0.7, respectively.}. Additionally, 
$\Delta \Phi_{\rm AB}$ depends on the position of both images and the source, and the lens 
potential at the image positions\cite{jack15,treu16}. Hence, the lensing mass distribution 
determines the value of the multiplicative factor $\Delta \Phi_{\rm AB}$.

We used a lens model to describe the primary lensing mass in SBS 0909+532 and SDSS J1339+1310. 
For each of these two double quasars, our lens model consisted of an elliptical surface mass 
density to account for the main lens galaxy G and an external shear $\gamma$ due to the 
gravitational action of other galaxies around the lens system. The surface mass density of G was 
modeled as a singular power-law distribution since a composite model (treating baryons and dark 
matter individually) leads to similar results\cite{suyu14,mill20}. In this preliminar study, 
instead of using high-resolution imaging to put constraints on the power-law index of G, we 
focused on an isothermal distribution, i.e., a singular isothermal ellipsoid (SIE). Such 
distribution is consistent with stellar and gas motions in the Milky Way, as well as 
observations of many spiral and elliptical galaxies. We also did not use the stellar kinematics 
of G\cite{para09}. 

We considered constraints on the time delay, the relative astrometry and the flux ratio between 
images, along with some observationally-motivated priors on SIE+$\gamma$ lens model parameters. 
These constraints/priors and the LENSMODEL software\cite{keet01} allowed us to simultaneously 
fit lens model parameters, position and flux of the source quasar, and $H_0^{\rm model}$ with 
dof = 0, where "dof" means degrees of freedom. In addition to the mass that is explicitly 
included in the lens model (main deflector plus external shear), we must take the mass along the 
line of sight to G into account. This additional effect can be approximated as an external 
convergence in the lens plane $\kappa_{\rm ext}$, which may be positive or negative depending on
the mass distribution along the sightline. The true time-delay distance $D_{\Delta t}^{\rm 
true}$ relates to that derived from the lens model and measured delay $D_{\Delta t}^{\rm model}$ 
by $D_{\Delta t}^{\rm true} = D_{\Delta t}^{\rm model}/(1 - \kappa_{\rm ext})$ (e.g., see Eq. 
(4) of Ref.~\citenum{wong20}), which leads to $H_0^{\rm true} = H_0^{\rm model}(1 - \kappa_{\rm 
ext})$. Therefore, when accounting for an external convergence, the Hubble constant 
decreases/increases in a factor $1 - \kappa_{\rm ext}$. The two next sections deal with 
estimates of $H_0^{\rm model}$ from observations of SBS 0909+532 and SDSS J1339+1310.                                                                                                           
                                                                        
\section{SBS 0909+532}\label{goico:sec3}
SBS 0909+532 is a doubly imaged quasar in which the background source (quasar) and the 
foreground early-type lens galaxy (main deflector) have redshifts $z_{\rm s}$ = 1.377 and 
$z_{\rm d}$ = 0.830, respectively\cite{koch97,osco97,lubi00}. Our first set of observational 
constraints consisted of the SBS 0909+532 time delay in \tref{goico:tbl1} taking a symmetric 
uncertainty (50 $\pm$ 3 days)\footnote{Despite 49 $\pm$ 3 days is fully consistent with the
measurement in \tref{goico:tbl1}, initially we have preferred to keep its central value and 
divide the error bar into two identical halves.}, the relative astrometry of B and G (their 
positions with respect to A at the origin of coordinates) and the flux of B in units such that 
the flux of A is equal to one. These last astro-photometric constraints were taken from the 
$HST$ near-IR data in Table 3 of Ref.~\citenum{leha00}. We also considered priors on the 
ellipticity $e$ and external shear of the SIE+$\gamma$ lens model described in 
\sref{goico:sec2}: $e \leq$ 0.5 (see Table 3 of Ref.~\citenum{slus12}) and $\gamma \leq$ 0.1 
(see Table 4 of Ref.~\citenum{leha00}).

\begin{figure}[h]
\begin{center}
\includegraphics[width=\textwidth]{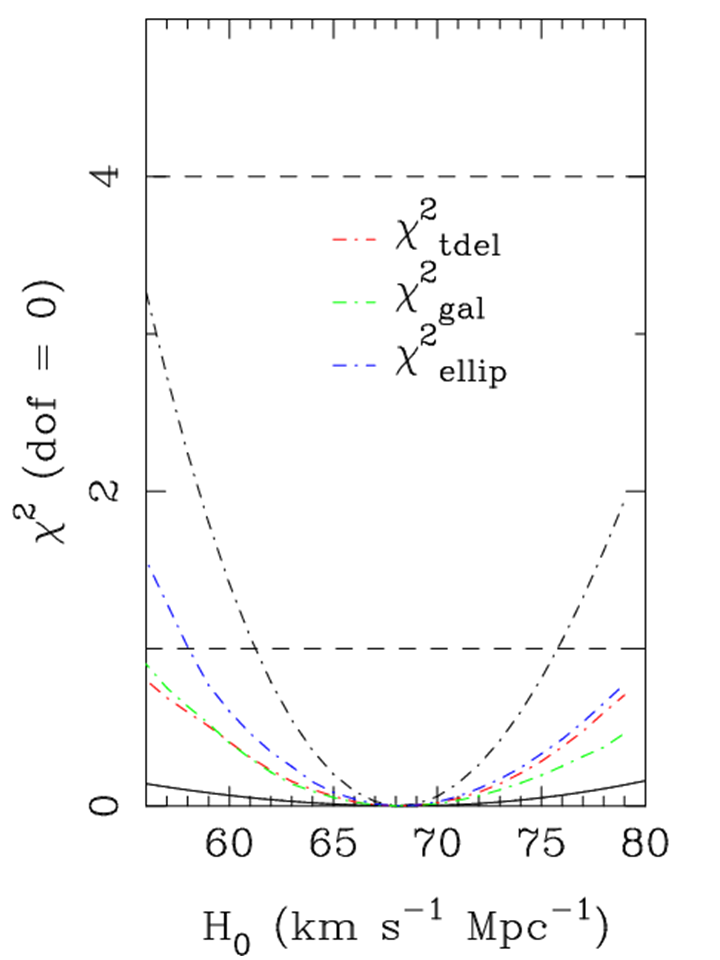}
\end{center}
\caption{Estimation of $H_0^{\rm model}$ from the SBS 0909+532 time delay and the 
astro-photometric constraints of Leh\'ar {\it et al.\/}\cite{leha00}. We also used 
observationally-motivated priors on the ellipticity of the lens galaxy and the external shear. 
We show the $\chi^2$ curve (black solid line) along with its 1$\sigma$ and 2$\sigma$ maximum 
thresholds (horizontal dashed lines). The black dashed-dotted line corresponds to an "improved" 
astrometry (see main text), with blue, green and red dashed-dotted lines describing some 
contributions to the total $\chi^2$.}
\label{goico:fig2}
\end{figure}

Although the data fit yielded a best solution for $H_0^{\rm model}$ of 68.4 (see 
\tref{goico:tbl2}), unfortunately, the $HST$ relative astrometry of Leh\'ar {\it et al.\/}\cite{leha00} 
is not so good as would be desirable. For instance, the relative position of the faint lens 
galaxy G was determined with a large uncertainty of about 100 mas (1 mas = 
0.$^{\prime\prime}$001). The insufficiently accurate astrometric measures were responsible for a 
broad valley in the $\chi^2$ curve (see the black solid line in \fref{goico:fig2}), so the 
1$\sigma$ confidence interval for $H_0^{\rm model}$ included values below 55 and above 80. If we 
were able to improve the Leh\'ar {\it et al.\/}'s astrometry, e.g., reducing errors in relative 
positions of B and G in factors 3 and 10, respectively, the best solution of $H_0^{\rm model}$ 
would be practically the same, but its uncertainty would be dramatically decreased to about 
10\%. Using "achievable" uncertainties of 1 mas in B and 10 mas in G, we obtained the black 
dashed-dotted line in \fref{goico:fig2} and $H_0^{\rm model}$ = 68.5 $\pm$ 7.5.

\begin{figure}[h]
\begin{center}
\includegraphics[width=\textwidth]{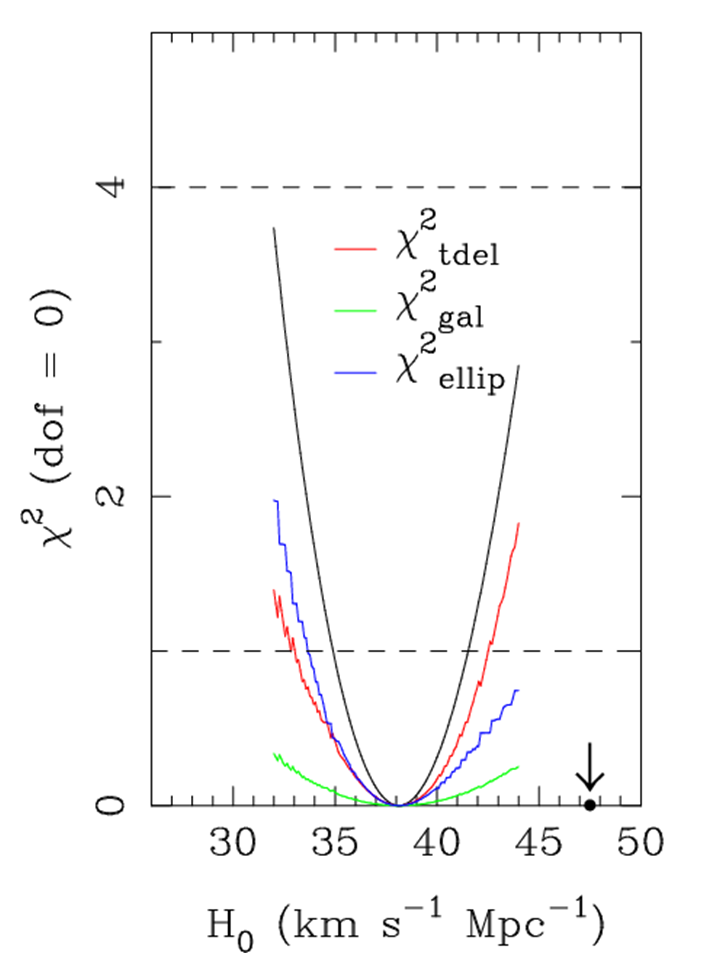}
\end{center}
\caption{Estimation of $H_0^{\rm model}$ from the SBS 0909+532 time delay and the 
astro-photometric constraints of Sluse {\it et al.\/}\cite{slus12}. The solid lines are related to 
priors on the shape of the lens galaxy (ellipticity and position angle; the black solid line 
represents the total $\chi^2$), while the black dot and vertical arrow indicate the best 
solution when using priors on the ellipticity and the external shear.}
\label{goico:fig3}
\end{figure}

In addition to new observations of the lens system, a reanalysis of the available $HST$ frames 
of SBS 0909+532 might produce a better astrometry for the system, and thus provide an accurate 
measure of the Hubble constant. This is a promising task that we and other astronomers are 
exploring. Sluse {\it et al.\/}\cite{slus12} have reanalysed the available $HST$ near-IR frames, 
obtaining a formally improved astrometry and even details on the structure of G. The error in 
the relative position of G was only 3 mas; about 30$-$40 times smaller than the uncertainty 
derived by Leh\'ar {\it et al.\/} We also considered these new constraints to measure $H_0^{\rm model}$. 
Using the SBS 0909+532 time delay with symmetric error (see above) and the astro-photometric 
solutions in Table 4 of Sluse {\it et al.\/}, along with the ellipticity and position angle of G in 
Table 3 of Sluse {\it et al.\/} (priors on the SIE+$\gamma$ lens model), we found $H_0^{\rm model}$ = 
38.2 $\pm$ 3.3 (see \tref{goico:tbl2} and the black solid line in \fref{goico:fig3}). Even the 
2$\sigma$ confidence interval only includes values below 45. Although a moderate increase in the 
best solution of $H_0^{\rm model}$ is found when taking the previous priors $e \leq$ 0.5 and 
$\gamma \leq$ 0.1 (see the black dot and vertical arrow in \fref{goico:fig3}), the Sluse {\it et al.\/}'s 
relative astrometry leads to best solutions below 50. Hence, either such astrometry is 
biased or the near-IR fluxes of the quasar images (optical emission) are strongly affected by 
microlensing in the lens galaxy\cite{medi05}.  

\begin{sidewaystable}
\tbl{Results for $H_0^{\rm model}$ using a SIE+$\gamma$ lens model (see main text).}
{\begin{tabular}{@{}cccccccccccccc@{}}
\toprule\\[-6pt]
 & \multicolumn{6}{c}{Observational constraints} &
 & \multicolumn{3}{c}{Priors on model parameters} &
 & \multicolumn{2}{c}{$H_0^{\rm model}$}\\[3pt]
\cline{2-7}\cline{9-11}\cline{13-14}\\[-6pt]
GLQ & $\Delta t_{\rm AB}$$^{\text a}$ & $\Delta x_{\rm AB}$$^{\text b}$ & $\Delta y_{\rm AB}$$^{\text b}$ 
& $\Delta x_{\rm AG}$$^{\text b}$ & $\Delta y_{\rm AG}$$^{\text b}$ & $F_{\rm B}/F_{\rm A}$$^{\text c}$ &
& $e$$^{\text d}$ & $\theta_e$$^{\text d}$ & $\gamma$$^{\text e}$ & 
& best$^{\text f}$ & 1$\sigma^{\text f}$\\[3.5pt]
\hline\\[-6pt]
SBS 0909+532 & 50 $\pm$ 3 & $-$0.987 $\pm$ 0.003 & $-$0.498 $\pm$ 0.003 & $-$0.415 $\pm$ 0.100 & 
$-$0.004 $\pm$ 0.100 & 0.89 $\pm$ 0.10 & & $\leq$ 0.5 & --- & $\leq$ 0.1 & & 68.4 & ---\\[3.5pt]
             &            & $-$0.987 $\pm$ 0.001 & $-$0.498 $\pm$ 0.001 & $-$0.415 $\pm$ 0.010 & 
$-$0.004 $\pm$ 0.010 &                 & &            &     &            & & 68.3 & 68.5 $\pm$ 7.5$^{\text g}$\\[3.5pt]  
             &            & $-$0.9868 $\pm$ 0.0006 & $-$0.4973 $\pm$ 0.0006 & $-$0.464 $\pm$ 0.003 & 
$-$0.055 $\pm$ 0.003 & 0.88 $\pm$ 0.10 & & 0.11 $\pm$ 0.08 & $-$48.1 $\pm$ 16.9 & --- & & 38.1 & 38.2 $\pm$ 3.3$^{\text h}$\\[3.5pt]
             &            &                      &                      &                      & 
             &            & & $\leq$ 0.5 & --- & $\leq$ 0.1 & & 47.5 & ---\\[3.5pt]
SDSS J1339+1310 & 47.0 $\pm$ 5.5 & +1.419 $\pm$ 0.001 & +0.939 $\pm$ 0.001 & +0.981 $\pm$ 0.010 & 
+0.485 $\pm$ 0.010 & 0.175 $\pm$ 0.015 & & 0.18 $\pm$ 0.05 & 32 $\pm$ 10 & ---- & & 69.1 & 69$^{+10}_{-8}$$^{\text i}$\\[3.5pt]
                & 48 $\pm$ 2 &           &     &    & 
&  & & & & & & 67.6 & 67.8 $\pm$ 4.4\\[3pt]
\Hline
\end{tabular}}
\begin{tabnote}
\\
$^{\text a}$ Time delay between both images in days. Some errors have been made symmetric.\\
$^{\text b}$ Relative positions of B and G with respect to A at the origin of coordinates. Here, $\Delta x$ 
and $\Delta y$ are given in arc seconds, and their positive directions are defined by west and north, 
respectively. For SBS 0909+532, some errors have been conveniently approximated.\\
$^{\text c}$ Flux ratio. For SBS 0909+532, errors are enlarged to 10\% to account for moderate microlensing 
effects.\\
$^{\text d}$ Ellipticity and position angle of the SIE. The position angle ($\theta_e$) is measured east of 
north.\\
$^{\text e}$ External shear strength.\\
$^{\text f}$ Best solution and 1$\sigma$ confidence interval for $H_0^{\rm model}$. We use standard units of 
km s$^{-1}$ Mpc$^{-1}$.\\
$^{\text g}$ Plausible but not real measurement. Astrometric errors have been reduced to "achievable" values
(see next row).\\
$^{\text h}$ Real measurement, but based on a biased astrometry or an inappropriate (strongly affected by 
microlensing) flux ratio.\\
$^{\text i}$ Measurement relying on an old, innacurate time delay.\\
\end{tabnote}
\label{goico:tbl2}
\end{sidewaystable}

\section{SDSS J1339+1310}\label{goico:sec4}
\begin{figure}[h]
\begin{center}
\includegraphics[width=\textwidth]{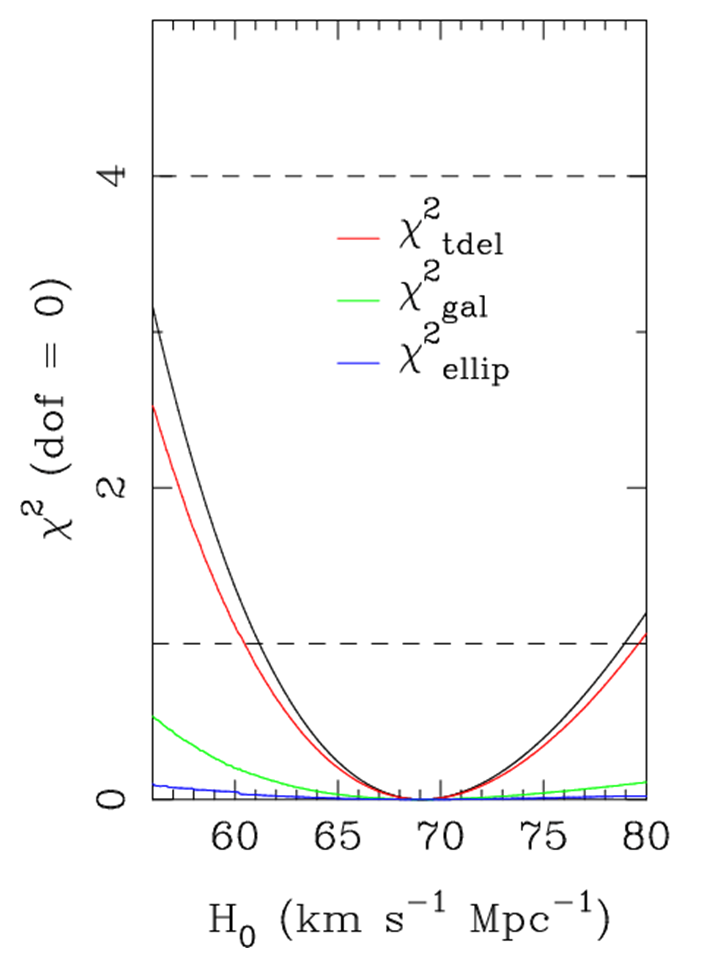}
\end{center}
\caption{Estimation of $H_0^{\rm model}$ from an old time delay of SDSS J1339+1310 with an 
accuracy of $\sim$12\% and constraints/priors from results in Refs.~\citenum{goic16} and 
\citenum{shal14}. The black line represents the total $\chi^2$, while the blue, green and red 
lines describe three different contributions to the total curve. The 1$\sigma$ and 2$\sigma$ 
maximum thresholds are also depicted (horizontal dashed lines).}
\label{goico:fig4}
\end{figure}

The gravitational lens system SDSS J1339+1310 was discovered by Inada {\it et al.\/}\cite{inad09}. It 
consists of two quasar images (A and B) at $z_{\rm s}$ = 2.231 and an early-type galaxy G at 
$z_{\rm d}$ = 0.607 acting as main deflector\cite{goic16}. The first set of observational 
constraints included the relative astrometry of B and G in the last column of Table 1 of 
Ref.~\citenum{shal14}, the macrolens magnification ratio from narrow-line/line-core flux ratios 
and a standard extinction law (based on emission lines in GTC spectra)\cite{goic16}, and an old
time delay from LT light curves\cite{goic16}. We note that the first time delay we used (47.0 
$\pm$ 5.5 days) is more inaccurate than the updated delay in \tref{goico:tbl1}. Additionally, we
have taken the ellipticity and position angle of G in the last column of Table 1 of Shalyapin et 
al.\cite{shal14} as priors on the SIE+$\gamma$ lens model. The data fit led to an 1$\sigma$ 
confidence interval $H_0^{\rm model}$ = 69$^{+10}_{-8}$ (accuracy of $\sim$13\%; see 
\tref{goico:tbl2} and the black line in \fref{goico:fig4}). The observational constraint on the 
time delay is the primary contribution to the $\chi^2$ curve (see the red line in 
\fref{goico:fig4}), while other constraints/priors (e.g., the position of G; see the green line 
in \fref{goico:fig4}) play a secondary role.

\begin{figure}[h]
\begin{center}
\includegraphics[width=\textwidth]{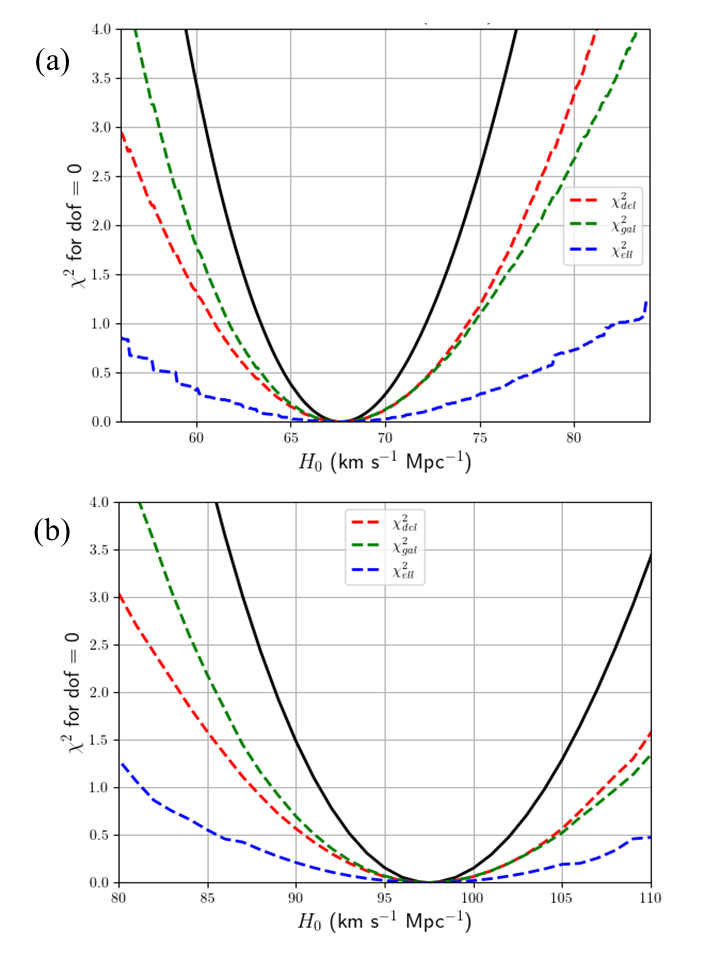}
\end{center}
\caption{Estimation of $H_0^{\rm model}$ from the updated time delay of SDSS J1339+1310 with an 
accuracy of $\sim$4\% and constraints/priors from results in Refs.~\citenum{goic16} and 
\citenum{shal14}. (a) SIE+$\gamma$ lens model. (b) DV+$\gamma$ lens model. To obtain
the $\chi^2$ curve in this bottom panel, we have assumed that light traces mass of the main lens 
galaxy (see main text).}
\label{goico:fig5}
\end{figure}    

Results in \fref{goico:fig4} suggest that a tighter constraint on the time delay would produce a
more accurate determination of the Hubble constant. Therefore, in a second approach, we used the 
updated time delay with a 4\% error that appears in \tref{goico:tbl1} to more accurate estimate 
$H_0^{\rm model}$. The new $\chi^2$ curve in \fref{goico:fig5}(a) indicates that $H_0^{\rm 
model}$ = 67.8 $\pm$ 4.4 (see also \tref{goico:tbl2}). This is a quite robust measurement of 
$H_0^{\rm model}$ because its relative error is small (only 6.5\%), and the priors on $e$ and 
$\theta_e$ do not play a relevant role (see the blue dashed line in \fref{goico:fig5}(a)). In 
addition, the macrolens magnification ratio is not affected by microlensing/extinction effects, 
and only one major issue must be addressed: the hypothesis about an isothermal mass distribution 
for G. Using 58 gravitational lens systems from the SLACS Survey, Koopmans {\it et al.\/}\cite{koop09} 
concluded that massive early-type galaxies have close to isothermal total density profiles, with 
a scatter between their logarithmic density slopes below 10\%. For a particular lens galaxy, a 
small deviation from the isothermal power-law index is plausible, and this potential deviation 
can be taken into account by increasing the error in $H_0^{\rm model}$ (see a more complete 
discussion in \sref{goico:sec5}). 

It is easy to demonstrate the need for dark matter (e.g., a power-law mass distribution) and the 
fact that a model in which light traces mass produces biased results. To this end, we again 
considered the updated time delay in \tref{goico:tbl1} and added a new prior, i.e., we worked 
with three priors instead two. Assuming that light traces mass of G, i.e., a de Vaucouleurs (DV) 
mass distribution instead a singular isothermal one, in a self-consistent way, the optical 
structure of G in the last column of Table 1 of Ref.~\citenum{shal14} (effective radius, 
ellipticity and position angle) was used to describe the structure of its mass. This scheme led 
to a biased $H_0^{\rm model}$ value of about 100 (97.7 $\pm$ 6.4; see \fref{goico:fig5}(b)). 
Even the 2$\sigma$ lower limit is above 85. 

\section{Discussion and future prospects}\label{goico:sec5}
Using two double quasars of the GLENDAMA sample (see \fref{goico:fig1}), we focused on the role 
that some observational constraints and hypotheses/priors on the mass model play in estimating 
$H_0^{\rm model}$ in a standard cosmology. The main lens galaxies in SBS 0909+532 and SDSS 
J1339+1310 were modelled with a singular isothermal ellipsoid, in agreement with observations in 
the Milky Way and SLACS Survey results for massive early-type galaxies acting as gravitational 
lenses\cite{koop09}. Adding the external shear $\gamma$ that is caused by galaxies around a lens 
system, we initially considered a SIE+$\gamma$ lens (mass) model. 

For SBS 0909+532, there are two different astrometric solutions based on the same $HST$ near-IR 
data. While the Leh\'ar {\it et al.\/}'s solution\cite{leha00} led to a best value of $H_0^{\rm model}$ 
equal to 68.4 and a broad 1$\sigma$ interval for this parameter, the Sluse {\it et al.\/}'s 
solution\cite{slus12} provided an 8.6\% measurement of $H_0^{\rm model}$ around a central value 
of 38.2 (we derived biased results making different choices of priors). Assuming that the time 
delay and flux ratio between quasar images that we used are right, the last astrometry would be 
biased. However, the observed near-IR fluxes correspond to optical emission from the quasar 
accretion disk, so they could be strongly affected by microlenses (stars) in the main lens 
galaxy\cite{medi05}. Hence, an accurate and reliable astrometric solution along with a detailed 
analysis of the macrolens magnification ratio (flux ratio free from extinction and microlensing 
effects) is required before robustly measuring $H_0^{\rm model}$ for a SIE+$\gamma$ scenario.     
    
Results for SDSS J1339+1310 are really encouraging because its current astrometry, updated time 
delay and macrolens magnification ratio through GTC spectra allowed us to accurately measure 
$H_0^{\rm model}$ (67.8 $\pm$ 4.4), with priors on the ellipticity and position angle of the SIE 
not playing a relevant role. It is also noteworthy that our 1$\sigma$ interval is not in tension 
with other recent estimates from GLQs\cite{wong20} and the CMB\cite{plan20} (see also 
Ref.~\citenum{free19}), and the central value practically coincides with the upper limit of the 
{\it Planck\/} collaboration. Accounting for a potential microlensing effect on the time 
delay\cite{tiek18} ($\sim$1 day) would only modify $H_0^{\rm model}$ by $\sim$2\%. Additionally, 
the use of a main galaxy mass model with an unrealistic density profile may have a significant 
impact on $H_0^{\rm model}$ and be responsible for an error of about 10\%\cite{schn13,koch20,
stac21}. At present, we do not know details about the mass density profile of the main deflector 
in SDSS J1339+1310, and thus we should adopt an uncertainty in $H_0^{\rm model}$ greater than 
that obtained with a SIE. Very recently, assuming that the deflectors of the H0LiCOW GLQs and 
the SLACS lenses share the same mass density properties, Birrer {\it et al.\/}\cite{birr20} have 
obtained $H_0$ = 67.4$^{+4.1}_{-3.2}$. This new GLQ-based result is in excellent agreement with 
ours and the CMB-based estimation of $H_0$, notably reduces tension between early and 
late-Universe probes, and illustrates the importance of assumptions on mass distributions. 

Future time-domain observations of large collections of GLQs will lead to robust constraints on 
$H_0$, and the matter and dark energy components of the Universe\cite{treu16}. The GLENDAMA 
project includes the first initiative to robotically monitor a small sample of 10 GLQs for about 
20 years\cite{gilm18}. This project and the associated robotic monitoring with the LT will end 
in 2025, after providing accurate time delays for several GLQs and discussing their cosmological 
implications. In next few years, other ongoing monitoring projects will also measure accurate 
delays for small/medium samples of GLQs (see the paper by Geoff Chih-Fan Chen in these 
proceedings), which will contribute to a rich database of tens of measured delays. Despite this 
optimistic perspective about time-domain results, some issues must be fixed before sheding light 
on unbiased values of cosmological parameters from such delay database. Deep spectroscopy, 
high-resolution imaging and other complementary observations will be required. For example, 
unaccounted mass along GLQ sightlines may produce overestimated/underestimated values of $H_0$ 
(see the end of \sref{goico:sec2}), so accurate $H_0$ estimates cannot ignore external 
convergences. Here, although we ignored the external convergence for SDSS J1339+1310, the 
unaccounted mass is expected to translate to a few percent relative uncertainty in 
$H_0$\cite{birr20,rusu17}, noticeably less than that related to the mass density profile of G 
(see above).  
 
\section*{Acknowledgments}
We thank the organizers of Sixteenth Marcel Grossmann Meeting for planning  a very interesting 
event and allowing us to give a talk in the parallel session "Cosmography with Gravitational 
Lensing". We also thank the chairs of such parallel session for creating a pleasant environment. 
We acknowledge Claudio Grillo, Mimoza Hafizi and Sherry Suyu for helpful comments that have 
significantly contributed to prepare the final text of this contribution. Among other things, 
the Gravitational LENses and DArk MAtter (GLENDAMA) project aims to construct accurate optical 
light curves of SBS 0909+532 and SDSS J1339+1310, and measure robust time delays for both 
systems. Although these optical variability studies mainly rely on observations with the 
Liverpool Telescope (LT), we are particularly grateful to our collaborators working in several 
institutions, who provide us with complementary data from the Maidanak Astronomical Observatory 
and the US Naval Observatory, and participate actively in the project development. The LT is 
operated on the island of La Palma by the Liverpool John Moores University (with financial 
support from the UK Science and Technology Facilities Council), in the Spanish Observatorio del 
Roque de los Muchachos of the Instituto de Astrof\'isica de Canarias. We thank the staff of the  
telescope for a kind interaction before, during and after the observations. This research has 
been supported by the grant AYA2017-89815-P funded by MCIN/AEI/10.13039/501100011033 and by 
“ERDF A way of making Europe”, and the grant PID2020-118990GB-I00 funded by 
MCIN/AEI/10.13039/501100011033.

\end{document}